%% file: template.tex
\documentclass{INTERSPEECH2023}
\usepackage{booktabs}
\usepackage{multirow}
\usepackage{enumitem}
\usepackage{caption}
\usepackage{hyperref}
\usepackage{listings}
\usepackage[linesnumbered, ruled]{algorithm2e}
\hypersetup{
    colorlinks=true,
    linkcolor=blue,
    filecolor=magenta,
    urlcolor=cyan,
    pdftitle={Overleaf Example},
    pdfpagemode=FullScreen,
    }

\usepackage{xcolor}

\definecolor{codegreen}{rgb}{0,0.6,0}
\definecolor{codegray}{rgb}{0.5,0.5,0.5}
\definecolor{codepurple}{rgb}{0.58,0,0.82}
\definecolor{backcolour}{rgb}{0.95,0.95,0.92}

\lstdefinestyle{mystyle}{
    backgroundcolor=\color{backcolour},
    commentstyle=\color{codegreen},
    keywordstyle=\color{magenta},
    numberstyle=\tiny\color{codegray},
    stringstyle=\color{codepurple},
    basicstyle=\ttfamily\scriptsize,
    breakatwhitespace=false,
    breaklines=true,
    captionpos=b,
    keepspaces=true,
    numbers=left,
    numbersep=5pt,
    showspaces=false,
    showstringspaces=false,
    showtabs=false,
    tabsize=2
}

\interspeechcameraready

\title{WhisperX: Time-Accurate Speech Transcription\\ of Long-Form Audio}
\name{Max Bain, Jaesung Huh, Tengda Han, Andrew Zisserman}
\vspace{-3mm}
\address{
  Visual Geometry Group, University of Oxford}
\email{maxbain@robots.ox.ac.uk}

\begin{document}

\twocolumn[{%
\renewcommand\twocolumn[1][]{#1}%
\maketitle
\vspace{-4em}
\begin{center}
    \centering
    \captionsetup{type=figure}
    \includegraphics[width=\textwidth]{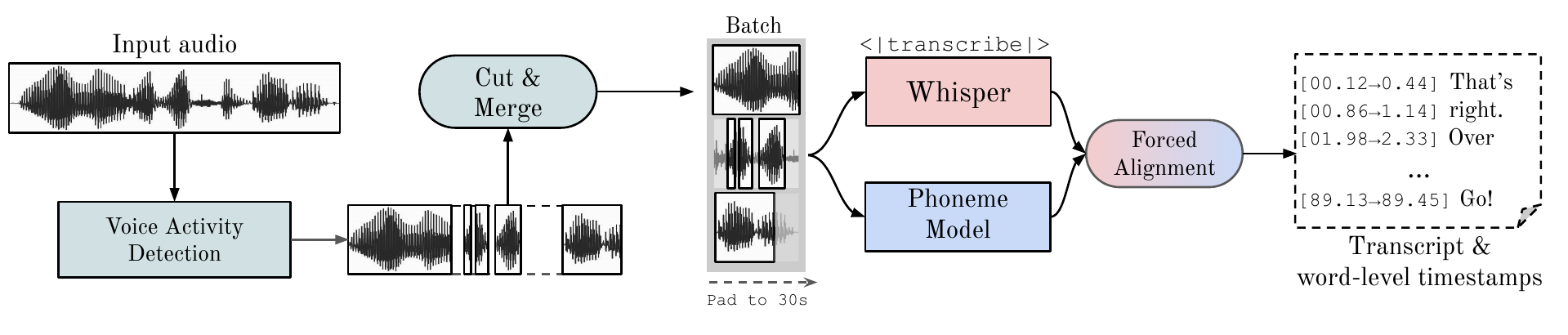}
    \captionof{figure}{\textbf{WhisperX}: We present a system for efficient speech transcription of long-form audio with \emph{word-level time alignment}. The input audio is first segmented with Voice Activity Detection and then cut \& merged into approximately 30-second input chunks with boundaries that lie on minimally active speech regions. The resulting chunks are then: (i) transcribed in parallel with Whisper, and (ii) forced aligned with a phoneme recognition model to produce accurate word-level timestamps at high throughput.}
\end{center}%
}]

\begin{abstract}
Large-scale, weakly-supervised speech recognition models, such as Whisper, have demonstrated impressive results on speech recognition across domains and languages.
However, the predicted timestamps corresponding to each utterance are prone to inaccuracies, and word-level timestamps are not available out-of-the-box. Further, their application to long audio via buffered transcription prohibits batched inference due to their sequential nature. To overcome the aforementioned challenges, we present \textit{WhisperX}, a time-accurate speech recognition system with word-level timestamps utilising voice activity detection and forced phoneme alignment. In doing so, we demonstrate state-of-the-art performance on long-form transcription and word segmentation benchmarks. Additionally, we show that pre-segmenting audio with our proposed VAD Cut \& Merge strategy improves transcription quality and enables a \textit{twelve-fold} transcription speedup via batched inference. The code is available open-source\footnote{\url{https://github.com/m-bain/whisperX}}.
\end{abstract}

\input{01_intro}

\input{02_method}

\input{03_results}

\section{Conclusion}
To conclude, we propose \textit{WhisperX}, a time-accurate speech recognition system enabling within-audio parallelised transcription. We show that the proposed VAD Cut \& Merge preprocessing reduces hallucination and repetition, enabling within-audio batched transcription, resulting in a twelve-fold speed increase without sacrificing transcription quality. Further, we show that the transcribed segments can be forced aligned with a phoneme model, providing accurate word-level segmentations with minimal inference overhead and resulting in time-accurate transcriptions benefitting a range of applications (e.g. subtitling, diarisation etc.).
A promising direction for future work is the training of a single-stage ASR system that can efficiently transcribe long-form audio with accurate word-level timestamps.
\\

\noindent\textbf{Acknowledgement}
This research is funded by the EPSRC VisualAI EP/T028572/1 (M. Bain, T. Han, A. Zisserman) and a Global Korea Scholarship (J. Huh).
Finally, the authors would like to thank the numerous open-source contributors and supporters of \textit{WhisperX}.

\bibliographystyle{IEEEtran}
\bibliography{egbib}

\end{document}

%% file: 01_intro.tex
\section{Introduction}

With the availability of large-scale web datasets, weakly-supervised and unsupervised training methods have demonstrated impressive performance on a multitude of speech processing tasks; including speech recognition~\cite{jia2019leveraging, baevski2020wav2vec, chen2022wavlm}, speaker recognition~\cite{kang2022augmentation,chen2023self}, speech separation~\cite{wisdom2020unsupervised}, and keyword spotting~\cite{gong2022ssast,prajwal2021visual}.
Whisper~\cite{radford2022robust} utilises this rich source of data to another scale. Leveraging 680,000 hours of noisy speech training data, including 96 other languages and 125,000 hours of English translation data, it showcases that weakly supervised pretraining of a simple encoder-decoder transformer~\cite{vaswani2017attention} can robustly achieve \textit{zero-shot} multilingual speech transcription on existing benchmarks.

Most of the academic benchmarks are comprised of short utterances, whereas real-world applications typically require transcribing long-form audio that can easily be hours or minutes long, such as meetings, podcasts and videos. Automatic Speech Recognition (ASR) models are typically trained on short audio segments (30 seconds for the case of Whisper) and the transformer architectures prohibit transcription of arbitrarily long input audio due to memory constraints.

Recent works~\cite{chiu2019comparison} employ heuristic sliding window style approaches that are prone to errors due to overlapping or incomplete audio (e.g. words being cut halfway through). Whisper proposes a buffered transcription approach that relies on accurate timestamp prediction to determine the amount to shift the subsequent input window by. Such a method is prone to severe drifting since timestamp inaccuracies in one window can accumulate to subsequent windows. The hand-crafted heuristics employed have achieved limited success.

A plethora of works exist on ``forced alignment'', aligning speech transcripts with audio at the word or phoneme level. Traditionally, this involves training acoustic phoneme models in a Hidden Markov Model (HMM)~\cite{brugnara1993automatic,gorman2011prosodylab, yuan2013automatic, mcauliffe2017montreal} framework using external boundary correction models~\cite{kim2002automatic, stolcke2014highly}.
Recent works employ deep learning strategies, such as a bi-directional attention matrix~\cite{li2022neufa} or CTC-segmentation with an end-to-end trained model~\cite{kurzinger2020ctc}. Further improvements may come from combining a state-of-the-art ASR model with a light-weight phoneme recognition model, both of which are trained with large-scale datasets.

To address these challenges, we propose \textit{WhisperX}, a system for efficient speech transcription of long-form audio with accurate word-level timestamps. It consists of three additional stages to Whisper transcription: (i) pre-segmenting the input audio with an external Voice Activity Detection (VAD) model; (ii) cut and merging the resulting VAD segments into approximately 30 seconds input chunks with boundaries lying on minimally active speech regions enabling batched whisper transcription; and finally (iii) forced alignment with an external phoneme model to provide accurate word-level timestamps.

%% file: 02_method.tex
\section{WhisperX}

In this section we describe \textit{WhisperX} and its components for long-form speech transcription with word-level alignment.

\subsection{Voice Activity Detection}
~\label{sec:vad_chunk}
Voice activity detection (VAD) refers to the process of identifying regions within an audio stream that contain speech. For \textit{WhisperX}, we first pre-segment the input audio with VAD. This provides the following three benefits: (1) VAD is much cheaper than ASR and avoids unnecessary forward passes of the latter during long inactive speech regions. (2) The audio can be sliced into chunks with boundaries that do not lie on active speech regions, thereby minimising errors due to boundary effects and enabling parallelised transcription. Finally, (3) the speech boundaries provided by the VAD model can be used to constrain the word-level alignment task to more local segments and remove reliance on Whisper timestamps -- which we show to be too unreliable.

VAD is typically formulated as a sequence labelling task; the input audio waveform is represented as a sequence of acoustic feature vectors extracted per time step $\mathbf{A} = \{a_1, a_2,..., a_T\}$ and the output is a sequence of binary labels $\mathbf{y} = \{y_1,y_2,...,y_T\}$, where $y_t = 1$ if there is speech at time step $t$ and $y_t=0$ otherwise.

In practice, the VAD model $\Omega_V: \mathbf{A} \rightarrow \mathbf{y}$ is instantiated as a neural network, whereby the output predictions $y_t \in [0,1]$ are post-processed with a \textit{binarize} step -- consisting of a smoothing stage (onset/offset thresholds) and decision stage (min. duration on/off)~\cite{8100927}.

The binary predictions can then represented as a sequence of active speech segments $\mathbf{s} = \{s_1,s_2,...,s_N\}$, with start and end indexes $s_i=(t_0^i,t_1^i)$.

\subsection{VAD Cut \& Merge}
Active speech segments $\mathbf{s}$ can be of arbitrary lengths, much shorter or longer than the maximum input duration of the ASR model, in this case Whisper. These Longer segments cannot be transcribed with a single forward pass. To address this, we propose a `min-cut' operation in the smoothing stage of the binary post-processing to provide an upper bound on the duration of active speech segments.

Specifically, we limit the length of active speech segments to be no longer than the maximum input duration of the ASR model. This is achieved by cutting longer speech segments at the point of minimum voice activation score (min-cut). To ensure the newly divided speech segments are not exceedingly short and have sufficient context, the cut is restricted between $\frac{1}{2}|\mathcal{A}_\text{train}|$ and $|\mathcal{A}_\text{train}|$, where $|\mathcal{A}_\text{train}|$ is the maximum duration of input audio during training (for Whisper this is 30 seconds). Pseudo-code detailing the proposed min-cut operation can be found in List.~\ref{lst:cut}.
\input{figures/fig-code_cut}

With an upper bound now set on the duration of input segments, the other extreme must be considered: very short segments, which present their distinct set of challenges. Transcribing brief speech segments eliminates the broader context beneficial for modelling speech in challenging scenarios. Moreover, transcribing numerous shorter segments increases total transcription time due to the increased number of forward passes required.

Therefore, we propose a `merge' operation, performed after `min-cut`, merging neighbouring segments with aggregate temporal spans less than a maximal duration threshold $\tau$ where $\tau \leq |\mathcal{A}_\text{train}|$. Empirically we find this to be optimal at $\tau = |\mathcal{A}_\text{train}|$, maximizing context during transcription and ensures the distribution of segment durations is closer to that observed during training.

\subsection{Whisper Transcription}
The resulting speech segments, now with duration approximately equal to the input size of the model, $|s_i| \approx |\mathcal{A}_\text{train}|~\forall i \in N$, and boundaries that do not lie on active speech, can be efficiently transcribed in parallel with Whisper $\Omega_{W}$, outputting text for each audio segment $\Omega_{W}:\mathbf{s}\rightarrow\mathcal{T}$.
We note that parallel transcription must be performed without conditioning on previous text, since the causal conditioning would otherwise break the independence assumption of each sample in the batch. In practice, we find this restriction to be beneficial, since conditioning on previous text is more prone to hallucination and repetition. We also use the no timestamp decoding method of Whisper.

\subsection{Forced Phoneme Alignment}

For each audio segment $s_i$ and its corresponding text transcription $\mathcal{T}_i$, consisting of a sequence of words $\mathcal{T}_i=[w_0, w_1, ..., w_m]$, our goal is to estimate the start and end time of each word. For this, we leverage a phoneme recognition model, trained to classify the smallest unit of speech distinguishing one word from another, \textit{e.g. the element p in ``tap"}. Let $\mathcal{C}$ be the set of phoneme classes in the model $\mathcal{C} = \{c_1,c_2,...,c_K\}$. Given an input audio segment, a phoneme classifier, takes an audio segment $S$ as input and outputs a logits matrix $L \in \mathbb{R}^{K \times T}$, where $T$ varies depending on the temporal resolution of the phoneme model.

Formally, for each segment, $s_i \in \mathbf{s}$, and its corresponding text $\mathcal{T}_i$: \textbf{(1)} Extract the unique set of phoneme classes in the segment text $\mathcal{T}_i$ common to the phoneme model, denoted by $\mathcal{C}_{\mathcal{T}_i} \subset \mathcal{C}$. \textbf{(2)} Perform phoneme classification over the input segment $s_i$, with the classification restricted to $\mathcal{C}_{\mathcal{T}_i}$ classes. \textbf{(3)} Apply Dynamic Time Warping (DTW) on the resulting logits matrix $L_i \in \mathbb{R}^{\mathcal{C}_{\mathcal{T}_i} \times T}$, to obtain the optimal temporal path of phonemes in $\mathcal{T}_i$. \textbf{(4)} Obtain start and end times for each word $w_i$ in $\mathcal{T}_i$ by taking the start and end time of the first and last phoneme within the word respectively.

For transcript phonemes not present in the phoneme model's dictionary $\mathcal{C}$, we assign the timestamp from the next nearest phoneme in the transcript. The \textit{for loop} described above can be batch processed in parallel, enabling fast transcription and word-alignment of long-form audio.

\subsection{Multi-lingual Transcription and Alignment}
\textit{WhisperX} can also be applied to multilingual transcription, with the caveat that (i) the VAD model should be robust to different languages, and (ii) the alignment phoneme model ought to be trained on the language(s) of interest. Multilingual phoneme recognition models~\cite{conneau2020unsupervised} are also a suitable option, possibly generalising to languages not seen during training -- this would just require an additional mapping from language-independent phonemes to the phonemes of the target language(s).\footnote{We were unable to find non-English ASR to evaluate multilingual word segmentation, but we show successful qualitative examples in the open-source repository.}

\subsection{Translation}
Whisper also offers a ``translate" mode that allows for translated transriptions from multiple languages into English. The batch VAD-based transcription can also be applied to the translation setting, however phoneme alignment is not possible due to there no longer being a phonetic audio-linguistic alignment between the speech and the translated transcript.

\subsection{Word-level Timestamps without Phoneme Recognition}

We explored the feasibility of extracting word-level timestamps from Whisper directly, without an external phoneme model, to remove the need for phoneme mapping and reduce inference overhead (in practice we find the alignment overhead is minimal, approx. $<$10\% in speed). Although attempts have been made to infer timestamps from cross-attention scores~\cite{lintoai2023whispertimestamped}, these methods under-perform when compared to our proposed external phoneme alignment approach, as evidenced in Section~\ref{sec:results}, and are prone to timestamp inaccuracies.

%% file: figures/fig-code_cut.tex
\begin{lstlisting}[language=Python, caption=Python code for the smoothing stage in the VAD binarize step\, with the proposed min\-cut operation to limit the length of segments., label=lst:cut]

def binarize_cut(scores,  max_dur,  onset_th,  offset_th, TIMESTEP):
  '''
  scores - array of VAD scores extracted at each TIMESTEP (e.g. 0.02 seconds)
  max_dur - maximum duration of ASR model
  onset_th - threshold for speech onset
  offset_th - threshold for speech offset
  
  returns:
  segs - array of active speech start and end
  '''
  segs = []
  start = 0
  is_active = scores[0] > offset_th
  max_len = int(max_dur * TIMESTEP)

  for i in range(1, len(scores)):
    sc = scores[t]
    if is_active:
      if i - start >= max_len:
        # min-cut modification
        pdx = i + max_len // 2
        qdx = i + max_len
        min_sp = argmin(scores[pdx:qdx])
        segs.append((start, pdx+min_sp))
        start = pdx + min_sp
      elif sc < offset_th:
        segs.append((start, i))
        is_active = False
    else:
      if sc > onset_th:
        start = i
        is_active = True

  return segs
\end{lstlisting}

%% file: 03_results.tex
\section{Evaluation}

Our evaluation addresses the following questions: (1) the effectiveness of \textit{WhisperX} for long-form transcription and word-level segmentation compared to state-of-the-art ASR models (namely Whisper and wav2vec2.0); (2) the benefit of VAD Cut \& Merge pre-processing in terms of transcription quality and speed; and (3) the effect of the choice of phoneme model and Whisper model on word segmentation performance.

\subsection{Datasets}

\noindent\textbf{The AMI Meeting Corpus.}
We used the test set of the AMI-IHM from the AMI Meeting Corpus~\cite{carletta2006ami} consisting of 16 audio recordings of meetings. Manually verified word-level alignments are provided for the test set used to evaluate word segmentation performance. \textbf{Switchboard-1 Telephone Corpus (SWB).} SWB~\cite{godfrey1993switchboard} consists of $\sim$2,400 hours of speech of telephone conversations. Ground truth transcriptions are provided with manually corrected word alignments. We randomly sub-sampled a set of 100 conversations. To evaluate long-form audio transcription, we report on \textbf{TEDLIUM-3}~\cite{hernandez2018ted} consisting of 11 TED talks, each 20 minutes in duration, and \textbf{Kincaid46}~\cite{kincaid46} consisting of various videos sourced from YouTube.

\subsection{Metrics}
For evaluating long-form audio transcription, we report word error rate (\textbf{WER}) and transcription speed (\textbf{Spd.}). To quantify the amount of repetition and hallucination, we measure insertion error rate (\textbf{IER}) and the number of 5-gram word duplicates within the predicted transcript (\textbf{5-Dup.}) respectively.
Since this does not evaluate the accuracy of the predicted timestamps, we also evaluate word segmentation metrics, for datasets that have word-level timestamps, jointly evaluating both transcription and timestamp quality. We report the Precision (\textbf{Prec.}) and Recall (\textbf{Rec.}) where a true positive is where a predicted word segment overlaps with a ground truth word segment within a collar, where both words are an exact string match. For all evaluations we use a collar value of 200 milliseconds to account for differences in annotation and models.

\begin{table}
\centering
\setlength{\tabcolsep}{5pt}
\footnotesize
\caption{Default configuration for WhisperX.}
\begin{tabular}{@{}lll@{}}
\toprule
\textbf{Type} & \textbf{Hyperparameter} & \textbf{Default Value} \\ \midrule
\multirow{5}{*}{VAD} & Model & \texttt{pyannote}~\cite{Bredin2020} \\
 & Onset threshold & 0.767 \\
 & Offset threshold & 0.377 \\
 & Min. duration on & 0.136 \\
 & Min. duration off & 0.067 \\ \midrule
\multirow{3}{*}{Whisper} & Model version & \texttt{large-v2} \\
 & Decoding strategy & greedy \\
 & Condition on previous text & False \\ \midrule
\multirow{3}{*}{\begin{tabular}[c]{@{}l@{}}Phoneme\\ Model\end{tabular}} & Architecture & wav2vec2.0 \\
 & Model version & \texttt{BASE\_960H} \\
 & Decoding strategy & greedy \\ \bottomrule
\end{tabular}
\label{tab:config}
\end{table}

\subsection{Implementation Details}
\noindent\textbf{WhisperX:} Unless specified otherwise, we use the default configuration in Table~\ref{tab:config} for all experiments. \textbf{Whisper~\cite{radford2022robust}:} For Whisper-only transcription and word-alignment we inherit the default configuration from Table~\ref{tab:config}, and use the official implementation\footnote{https://github.com/openai/whisper/releases/tag/v20230307} for inferring word timestamps. \textbf{Wav2vec2.0~\cite{baevski2020wav2vec}:} For wav2vec2.0 transcription and word-alignment we use the default settings in Table~\ref{tab:config} unless specified otherwise. We obtain the various model versions from the official torchaudio repository\footnote{https://pytorch.org/audio/stable/pipelines.html\#module-torchaudio.pipelines} and build upon the forced alignment tutorial~\cite{yang2022torchaudio}. Base\_960h and Large\_960h models were trained on Librispeech~\cite{panayotov2015librispeech} data, whereas the VoxPopuli model was trained on the Voxpopuli~\cite{wang2021voxpopuli} corpus.
For benchmarking inference speed, all models are measured on an NVIDIA A40 gpu, as multiples of Whisper's speed.

\input{tables/sota}

\subsection{Results}
\label{sec:results}

\input{tables/vad_batching}

\subsubsection{Word Segmentation Performance}
Comparing to previous state-of-the-art speech transcription models (Table~\ref{tab:sota}), Whisper and wav2vec2.0, we find that \textit{WhisperX} substantially outperforms both in word segmentation benchmarks, WER, and transcription speed. Especially with batched transcription, \textit{WhisperX} even surpasses the speed of the lightweight wav2vec2 model. However, solely using Whisper for word-level timestamps extraction significantly underperforms in word segmentation precision and recall on both SWB and AMI corpuses, even falling short of wav2vec2.0, a smaller model with less training data. This implies the insufficiency of Whisper's large-scale noisy training data and current architecture for learning accurate word-level timestamps.

\subsubsection{Effect of VAD Chunking}
Table~\ref{tab:vad_batch} demonstrates the benefits of pre-segmenting audio with VAD and Cut \& Merge operations, improving both transcription-only WER and word segmentation precision and recall. Batched transcription without VAD chunking, however, degrades both transcription quality and word segmentation due to boundary effects.

Batched inference with VAD, transcribing each segment independently, provides a nearly twelve-fold speed increase without performance loss, overcoming the limitations of buffered transcription~\cite{radford2022robust}. Batch inference without VAD, using a sliding window, significantly degrades WER due to boundary effects, even with heuristic overlapped chunking as in huggingface\footnote{https://huggingface.co/openai/whisper-large}.

The optimal merge threshold value for Cut \& Merge operations $\tau$ is found to be the input duration that Whisper was trained on $|\mathcal{A}_\text{train}| = 30$, which provides the fastest transcription speed and lowest WER. This confirms that maximum context yields the most accurate transcription.

\subsubsection{Hallucination \& Repetition}
In Table~\ref{tab:sota}, we find that \textit{WhisperX} reports the lowest IER on the Kincaid46 and TED-LIUM benchmarks, confirming that the proposed VAD Cut \& Merge operations reduce hallucination in Whisper. Further, we find that repetition errors, measuring by counting the total number of 5-gram duplicates per audio, is also reduced by the proposed VAD operations. By removing the reliance on decoded timestamp tokens, and instead using external VAD segment boundaries, \textit{WhisperX} avoids repetitive transcription loops and hallucinating speech during inactivate speech regions.

Whilst wav2vec2.0 underperforms in both WER and word segmentation, we find that it is far less prone to repetition errors compared to both Whisper and \textit{WhisperX}. Further work is needed to reduce hallucination and repetition errors.

\input{tables/arch_ablation}

\subsubsection{Effect of Chosen Whisper and Alignment Models}
We compare the effect of different Whisper and phoneme recognition models on word segmentation performance across the AMI and SWB corpuses in Table~\ref{tab:arch_ablation}. Unsurprisingly, we see consistent improvements in both precision and recall when using a larger Whisper model.
In contrast, the bigger phoneme model is not necessarily the best and the results are more nuanced. The model trained on the VoxPopuli corpus significantly outperforms other models on AMI, suggesting that there is a higher degree of domain similarity between the two corpora.

The large alignment model does not show consistent gains, suggesting the need for additional supervised training data. Overall the base model trained on LibriSpeech performs consistently well and should be the default alignment model for \textit{WhisperX}.

%% file: tables/sota.tex
\begin{table*}[!ht]
\centering
\footnotesize
\setlength{\tabcolsep}{8pt}
\caption{\textbf{State-of-the-art comparison of long-form audio transcription and word segmentation} on the TED-LIUM, Kincaid46, AMI, and SWB corpora. \textbf{Spd} denotes transcription speed, \textbf{WER} denotes Word Error Rate, 
\textbf{5-Dup} denotes the \textnumero~5-gram duplicates, Precision \& Recall are calculated with a collar value of 200ms. \dag Word timestamps from Whisper are not directly available but are inferred via Dynamic Time Warping of the decoded tokens attention scores.}
\begin{tabular}{@{}lrrrrrrrrrrr@{}}
\toprule
\multicolumn{1}{c}{\multirow{2}{*}{Model}} & \multicolumn{4}{c}{TED-LIUM~\cite{hernandez2018ted}} & \multicolumn{3}{c}{Kincaid46~\cite{kincaid46}} & \multicolumn{2}{c}{AMI~\cite{carletta2006ami}} & \multicolumn{2}{c}{SWB~\cite{godfrey1993switchboard}} \\ \cmidrule(l){2-12} 
\multicolumn{1}{c}{} & \multicolumn{1}{c}{Spd.$\uparrow$} & \multicolumn{1}{c}{WER$\downarrow$} & \multicolumn{1}{c}{IER$\downarrow$} & \multicolumn{1}{c}{5-Dup.$\downarrow$} & \multicolumn{1}{c}{WER$\downarrow$} & \multicolumn{1}{c}{IER$\downarrow$} & \multicolumn{1}{c}{5-Dup.$\downarrow$} & \multicolumn{1}{c}{Prec.$\uparrow$} & \multicolumn{1}{c}{Rec.$\uparrow$} & \multicolumn{1}{c}{Prec.$\uparrow$} & \multicolumn{1}{c}{Rec.$\uparrow$} \\ \midrule
wav2vec2.0~\cite{baevski2020wav2vec} & 10.3$\times$ & 19.8 & 8.5 & \textbf{129} &  
28.0 &  5.3 &\textbf{ 29} & 81.8 & 45.5 & 92.9 & 54.3 \\
Whisper~\cite{radford2022robust} & 1.0$\times$ & 10.5 & 7.7 & 221 & 12.5 & 3.2 & 131 & 78.9 & 52.1 & 85.4 & 62.8 \\ \midrule
\textbf{WhisperX} & \textbf{11.8$\times$} & \textbf{9.7} & \textbf{6.7} & 189 & \textbf{11.8} & \textbf{2.2} & 75 & \textbf{84.1} & \textbf{60.3} & \textbf{93.2} & \textbf{65.4} \\ \bottomrule
\end{tabular}
\label{tab:sota}
\vspace{-0.5em}
\end{table*}

%% file: tables/vad_batching.tex
\begin{table}
\centering
\setlength{\tabcolsep}{7pt}
\footnotesize
\caption{\textbf{Effect of WhisperX's VAD Cut \& Merge and batched transcription on long-form audio transcription} on the TED-LIUM benchmark and AMI corpus. Full audio input corresponds to WhisperX without any VAD pre-processing, VAD-CM$_{\tau}$ refers to VAD pre-processing with Cut \& Merge, where $\tau$ is the merge duration threshold in seconds.}
\begin{tabular}{@{}lcrrrr@{}}
\toprule
\multicolumn{1}{c}{\multirow{2}{*}{\textbf{Input}}} & \multirow{2}{*}{\textbf{\begin{tabular}[c]{@{}c@{}}Batch\\ Size \end{tabular}}} & \multicolumn{2}{c}{\textbf{TED-LIUM}} & \multicolumn{2}{c}{\textbf{AMI}} \\ \cmidrule(l){3-6} 
\multicolumn{1}{c}{} &  & \multicolumn{1}{c}{\textbf{WER$\downarrow$}} & \multicolumn{1}{c}{\textbf{Spd.$\uparrow$}} & \multicolumn{1}{c}{\textbf{Prec.$\uparrow$}} & \multicolumn{1}{c}{\textbf{Rec.$\uparrow$}} \\ \midrule
\multirow{2}{*}{Full audio} & 1 & 10.52 & 1.0$\times$ & 82.6 & 53.4 \\
 & 32 &  78.78 & 7.1$\times$ & 43.2 & 25.7 \\
 \midrule
\multirow{2}{*}{VAD-CM$_{15}$} & 1 &  \multirow{2}{*}{9.72} & 2.1$\times$ & \multirow{2}{*}{84.1} & \multirow{2}{*}{56.0} \\
 & 32 &  & 7.9$\times$ &  & \\ \midrule
  \multirow{2}{*}{VAD-CM$_{30}$} & 1 & \multirow{2}{*}{\textbf{9.70}} & 2.7$\times$ & \multirow{2}{*}{\textbf{84.1}} & \multirow{2}{*}{\textbf{60.3}} \\
 & 32 & & \textbf{11.8$\times$} &  & \\ \bottomrule
\end{tabular}
\vspace{-0.5em}
\label{tab:vad_batch}
\end{table}

%% file: tables/arch_ablation.tex
\begin{table}
\setlength{\tabcolsep}{7pt}
\centering
\footnotesize
\caption{\textbf{Effect of whisper model and phoneme model on WhisperX on word segmentation.} Both the choice of whisper and phoneme model has a significant effect on word segmentation performance.}
\begin{tabular}{@{}llrrrr@{}}
\toprule
\multicolumn{1}{c}{\multirow{2}{*}{\textbf{\begin{tabular}[c]{@{}c@{}}Whisper\\ Model\end{tabular}}}} & \multicolumn{1}{c}{\multirow{2}{*}{\textbf{\begin{tabular}[c]{@{}c@{}}Phoneme\\ Model\end{tabular}}}} & \multicolumn{2}{c}{\textbf{AMI}} & \multicolumn{2}{c}{\textbf{SWB}} \\ \cmidrule(l){3-6} 
\multicolumn{1}{c}{} & \multicolumn{1}{c}{} & \multicolumn{1}{c}{\textbf{Prec.}} & \multicolumn{1}{c}{\textbf{Rec.}} & \multicolumn{1}{c}{\textbf{Prec.}} & \multicolumn{1}{c}{\textbf{Rec.}} \\ \midrule
\multirow{3}{*}{base.en} & Base\_960h & 83.7 & 58.9 & \textbf{93.1} & \textbf{64.5} \\
 & Large\_960h & 84.9 & 56.6 & \textbf{93.1} & 62.9 \\
 & VoxPopuli & \textbf{87.4} & \textbf{60.3} & 86.3 & 60.1 \\ \midrule
\multirow{3}{*}{small.en} & Base\_960h & 84.1 & 59.4 & 92.9 & 62.7 \\
  & Large\_960h & 84.6 & 55.7 & \textbf{94.0} &\textbf{ 64.9} \\
 & VoxPopuli & \textbf{87.7} & \textbf{61.2} & 84.7 & 56.3 \\ \midrule
\multirow{3}{*}{large-v2} & Base\_960h & 84.1 & 60.3 & 93.2 & 65.4 \\
  & Large\_960h & 84.9 & 57.1 & \textbf{93.5} & \textbf{65.7} \\
 & VoxPopuli & \textbf{87.7} & \textbf{61.7} & 84.9 & 58.7 \\ \bottomrule
\end{tabular}
\label{tab:arch_ablation}
\vspace{-0.5em}
\end{table}

%% file: template.bbl
\begin{thebibliography}{10}
\providecommand{\url}[1]{#1}
\csname url@samestyle\endcsname
\providecommand{\newblock}{\relax}
\providecommand{\bibinfo}[2]{#2}
\providecommand{\BIBentrySTDinterwordspacing}{\spaceskip=0pt\relax}
\providecommand{\BIBentryALTinterwordstretchfactor}{4}
\providecommand{\BIBentryALTinterwordspacing}{\spaceskip=\fontdimen2\font plus
\BIBentryALTinterwordstretchfactor\fontdimen3\font minus
  \fontdimen4\font\relax}
\providecommand{\BIBforeignlanguage}[2]{{%
\expandafter\ifx\csname l@#1\endcsname\relax
\typeout{** WARNING: IEEEtran.bst: No hyphenation pattern has been}%
\typeout{** loaded for the language `#1'. Using the pattern for}%
\typeout{** the default language instead.}%
\else
\language=\csname l@#1\endcsname
\fi
#2}}
\providecommand{\BIBdecl}{\relax}
\BIBdecl

\bibitem{jia2019leveraging}
Y.~Jia, M.~Johnson, W.~Macherey, R.~J. Weiss, Y.~Cao, C.-C. Chiu, N.~Ari,
  S.~Laurenzo, and Y.~Wu, ``Leveraging weakly supervised data to improve
  end-to-end speech-to-text translation,'' in \emph{ICASSP}.\hskip 1em plus
  0.5em minus 0.4em\relax IEEE, 2019, pp. 7180--7184.

\bibitem{baevski2020wav2vec}
A.~Baevski, Y.~Zhou, A.~Mohamed, and M.~Auli, ``wav2vec 2.0: A framework for
  self-supervised learning of speech representations,'' \emph{NeurIPS},
  vol.~33, pp. 12\,449--12\,460, 2020.

\bibitem{chen2022wavlm}
S.~Chen, C.~Wang, Z.~Chen, Y.~Wu, S.~Liu, Z.~Chen, J.~Li, N.~Kanda,
  T.~Yoshioka, X.~Xiao \emph{et~al.}, ``Wavlm: Large-scale self-supervised
  pre-training for full stack speech processing,'' \emph{IEEE Journal of
  Selected Topics in Signal Processing}, vol.~16, no.~6, pp. 1505--1518, 2022.

\bibitem{kang2022augmentation}
J.~Kang, J.~Huh, H.~S. Heo, and J.~S. Chung, ``Augmentation adversarial
  training for self-supervised speaker representation learning,'' \emph{IEEE
  Journal of Selected Topics in Signal Processing}, vol.~16, no.~6, pp.
  1253--1262, 2022.

\bibitem{chen2023self}
H.~Chen, H.~Zhang, L.~Wang, K.~A. Lee, M.~Liu, and J.~Dang, ``Self-supervised
  audio-visual speaker representation with co-meta learning,'' in
  \emph{ICASSP}, 2023.

\bibitem{wisdom2020unsupervised}
S.~Wisdom, E.~Tzinis, H.~Erdogan, R.~Weiss, K.~Wilson, and J.~Hershey,
  ``Unsupervised sound separation using mixture invariant training,''
  \emph{NeurIPS}, vol.~33, pp. 3846--3857, 2020.

\bibitem{gong2022ssast}
Y.~Gong, C.-I. Lai, Y.-A. Chung, and J.~Glass, ``Ssast: Self-supervised audio
  spectrogram transformer,'' in \emph{Proc. AAAI}, vol.~36, no.~10, 2022, pp.
  10\,699--10\,709.

\bibitem{prajwal2021visual}
K.~Prajwal, L.~Momeni, T.~Afouras, and A.~Zisserman, ``Visual keyword spotting
  with attention,'' \emph{arXiv preprint arXiv:2110.15957}, 2021.

\bibitem{radford2022robust}
A.~Radford, J.~W. Kim, T.~Xu, G.~Brockman, C.~McLeavey, and I.~Sutskever,
  ``Robust speech recognition via large-scale weak supervision,'' \emph{arXiv
  preprint arXiv:2212.04356}, 2022.

\bibitem{vaswani2017attention}
A.~Vaswani, N.~Shazeer, N.~Parmar, J.~Uszkoreit, L.~Jones, A.~N. Gomez,
  {\L}.~Kaiser, and I.~Polosukhin, ``Attention is all you need,''
  \emph{NeurIPS}, vol.~30, 2017.

\bibitem{chiu2019comparison}
C.-C. Chiu, W.~Han, Y.~Zhang, R.~Pang, S.~Kishchenko, P.~Nguyen, A.~Narayanan,
  H.~Liao, S.~Zhang, A.~Kannan \emph{et~al.}, ``A comparison of end-to-end
  models for long-form speech recognition,'' in \emph{2019 IEEE automatic
  speech recognition and understanding workshop (ASRU)}.\hskip 1em plus 0.5em
  minus 0.4em\relax IEEE, 2019, pp. 889--896.

\bibitem{brugnara1993automatic}
F.~Brugnara, D.~Falavigna, and M.~Omologo, ``Automatic segmentation and
  labeling of speech based on hidden markov models,'' \emph{Speech
  Communication}, vol.~12, no.~4, pp. 357--370, 1993.

\bibitem{gorman2011prosodylab}
K.~Gorman, J.~Howell, and M.~Wagner, ``Prosodylab-aligner: A tool for forced
  alignment of laboratory speech,'' \emph{Canadian Acoustics}, vol.~39, no.~3,
  pp. 192--193, 2011.

\bibitem{yuan2013automatic}
J.~Yuan, N.~Ryant, M.~Liberman, A.~Stolcke, V.~Mitra, and W.~Wang, ``Automatic
  phonetic segmentation using boundary models.'' in \emph{Interspeech}, 2013,
  pp. 2306--2310.

\bibitem{mcauliffe2017montreal}
M.~McAuliffe, M.~Socolof, S.~Mihuc, M.~Wagner, and M.~Sonderegger, ``Montreal
  forced aligner: Trainable text-speech alignment using kaldi.'' in
  \emph{Interspeech}, vol. 2017, 2017, pp. 498--502.

\bibitem{kim2002automatic}
Y.-J. Kim and A.~Conkie, ``Automatic segmentation combining an hmm-based
  approach and spectral boundary correction,'' in \emph{Seventh International
  conference on spoken language processing}, 2002.

\bibitem{stolcke2014highly}
A.~Stolcke, N.~Ryant, V.~Mitra, J.~Yuan, W.~Wang, and M.~Liberman, ``Highly
  accurate phonetic segmentation using boundary correction models and system
  fusion,'' in \emph{ICASSP}.\hskip 1em plus 0.5em minus 0.4em\relax IEEE,
  2014, pp. 5552--5556.

\bibitem{li2022neufa}
J.~Li, Y.~Meng, Z.~Wu, H.~Meng, Q.~Tian, Y.~Wang, and Y.~Wang, ``Neufa: Neural
  network based end-to-end forced alignment with bidirectional attention
  mechanism,'' in \emph{ICASSP}.\hskip 1em plus 0.5em minus 0.4em\relax IEEE,
  2022, pp. 8007--8011.

\bibitem{kurzinger2020ctc}
L.~K{\"u}rzinger, D.~Winkelbauer, L.~Li, T.~Watzel, and G.~Rigoll,
  ``Ctc-segmentation of large corpora for german end-to-end speech
  recognition,'' in \emph{Speech and Computer (SPECOM 2020)}.\hskip 1em plus
  0.5em minus 0.4em\relax Springer, 2020, pp. 267--278.

\bibitem{8100927}
G.~Gelly and J.-L. Gauvain, ``Optimization of rnn-based speech activity
  detection,'' \emph{IEEE/ACM Transactions on Audio, Speech, and Language
  Processing}, vol.~26, no.~3, pp. 646--656, 2018.

\bibitem{conneau2020unsupervised}
A.~Conneau, A.~Baevski, R.~Collobert, A.~Mohamed, and M.~Auli, ``Unsupervised
  cross-lingual representation learning for speech recognition,'' \emph{arXiv
  preprint arXiv:2006.13979}, 2020.

\bibitem{lintoai2023whispertimestamped}
J.~Louradour, ``whisper-timestamped,''
  \url{https://github.com/linto-ai/whisper-timestamped/tree/f861b2b19d158f3cbf4ce524f22c78cb471d6131},
  2023.

\bibitem{carletta2006ami}
J.~Carletta, S.~Ashby, S.~Bourban, M.~Flynn, M.~Guillemot, T.~Hain, J.~Kadlec,
  V.~Karaiskos, W.~Kraaij, M.~Kronenthal \emph{et~al.}, ``The ami meeting
  corpus: A pre-announcement,'' in \emph{Machine Learning for Multimodal
  Interaction: Second International Workshop, MLMI 2005, Edinburgh, UK, July
  11-13, 2005, Revised Selected Papers 2}.\hskip 1em plus 0.5em minus
  0.4em\relax Springer, 2006, pp. 28--39.

\bibitem{godfrey1993switchboard}
J.~Godfrey and E.~Holliman, ``Switchboard-1 release 2 ldc97s62,''
  \emph{Linguistic Data Consortium}, p.~34, 1993.

\bibitem{hernandez2018ted}
F.~Hernandez, V.~Nguyen, S.~Ghannay, N.~Tomashenko, and Y.~Esteve, ``Ted-lium
  3: Twice as much data and corpus repartition for experiments on speaker
  adaptation,'' in \emph{Speech and Computer: 20th International Conference,
  SPECOM 2018, Leipzig, Germany, September 18--22, 2018, Proceedings 20}.\hskip
  1em plus 0.5em minus 0.4em\relax Springer, 2018, pp. 198--208.

\bibitem{kincaid46}
``Which automatic transcription service is the most accurate?''
  https://medium.com/descript/which-automatic-transcription-service-is-the-most-accurate-2018-2e859b23ed19,
  accessed: 2023-04-27.

\bibitem{Bredin2020}
H.~{Bredin}, R.~{Yin}, J.~M. {Coria}, G.~{Gelly}, P.~{Korshunov},
  M.~{Lavechin}, D.~{Fustes}, H.~{Titeux}, W.~{Bouaziz}, and M.-P. {Gill},
  ``{pyannote.audio: neural building blocks for speaker diarization},'' in
  \emph{ICASSP}, 2020.

\bibitem{yang2022torchaudio}
Y.-Y. Yang, M.~Hira, Z.~Ni, A.~Astafurov, C.~Chen, C.~Puhrsch, D.~Pollack,
  D.~Genzel, D.~Greenberg, E.~Z. Yang \emph{et~al.}, ``Torchaudio: Building
  blocks for audio and speech processing,'' in \emph{ICASSP 2022-2022 IEEE
  International Conference on Acoustics, Speech and Signal Processing
  (ICASSP)}.\hskip 1em plus 0.5em minus 0.4em\relax IEEE, 2022, pp. 6982--6986.

\bibitem{panayotov2015librispeech}
V.~Panayotov, G.~Chen, D.~Povey, and S.~Khudanpur, ``Librispeech: an asr corpus
  based on public domain audio books,'' in \emph{2015 IEEE international
  conference on acoustics, speech and signal processing (ICASSP)}.\hskip 1em
  plus 0.5em minus 0.4em\relax IEEE, 2015, pp. 5206--5210.

\bibitem{wang2021voxpopuli}
C.~Wang, M.~Riviere, A.~Lee, A.~Wu, C.~Talnikar, D.~Haziza, M.~Williamson,
  J.~Pino, and E.~Dupoux, ``Voxpopuli: A large-scale multilingual speech corpus
  for representation learning, semi-supervised learning and interpretation,''
  \emph{arXiv preprint arXiv:2101.00390}, 2021.

\end{thebibliography}
